\documentclass[aps,prb,twocolumn,superscriptaddress,showpacs]{revtex4-1}
\usepackage{graphicx,color,ulem}

\definecolor{red}{rgb}{1,0,0}

\definecolor{green}{rgb}{0,1,0}

\definecolor{blue}{rgb}{0,0,1}

\begin{document}

\title{Evolution of the density of states at the Fermi level across the metal-to-insulator crossover in alkali doped zeolite}

\author{Mutsuo Igarashi}
\email{e-mail: igarashi@elc.gunma-ct.ac.jp}
\affiliation{Laboratory for Applied Physics, Department of Electrical Engineering, Gunma College, National Institute of Technology, Toribamachi 580, Maebashi 371-8530, Gunma, Japan}

\author{Peter Jegli\v{c}}
\email{e-mail: peter.jeglic@ijs.si}
\affiliation{Jo\v{z}ef Stefan Institute, Jamova 39, 1000 Ljubljana, Slovenia}
\affiliation{EN-FIST Centre of Excellence, Dunajska 156, 1000 Ljubljana,
Slovenia}

\author{Andra\v{z} Krajnc}
\affiliation{National Institute of Chemistry, Hajdrihova 19, 1001 Ljubljana, Slovenia}

\author{Rok \v{Z}itko}
\affiliation{Jo\v{z}ef Stefan Institute, Jamova 39, 1000 Ljubljana, Slovenia}

\author{Takehito Nakano}
\affiliation{Department of Physics, Graduate School of Science, Osaka University, Toyonaka 560-0043, Osaka, Japan}

\author{Yasuo Nozue}
\affiliation{Department of Physics, Graduate School of Science, Osaka University, Toyonaka 560-0043, Osaka, Japan}

\author{Denis Ar\v{c}on}
\affiliation{Jo\v{z}ef Stefan Institute, Jamova 39, 1000 Ljubljana, Slovenia}
\affiliation{Faculty of Mathematics and Physics, University of Ljubljana, Jadranska 19, 1000 Ljubljana, Slovenia}
\date{\today}

\begin{abstract}
We report a systematic nuclear magnetic resonance investigation of the $^{23}$Na spin-lattice relaxation rate, $1/T_1$, in sodium loaded low-silica X (LSX) zeolite, Na$_n$/Na$_{12}$-LSX, for various loading levels of sodium atoms $n$ across the metal-to-insulator crossover. 
For high loading levels of $n \geq 14.2$, $1/T_1T$ shows nearly temperature-independent behavior between 10 K and 25 K consistent with the Korringa relaxation mechanism and metallic ground state.
As the loading levels decrease below $n \leq 11.6$, the extracted density of states (DOS) at the Fermi level sharply decreases, although a residual DOS at Fermi level is still observed even in samples that lack the metallic Drude-peak in the optical reflectance.
The observed crossover is a result of a complex loading-level dependence of electric potential felt by the electrons confined to zeolite cages, where the electronic correlations and disorder both play an important role.
\end{abstract}

\maketitle

\section{Introduction}

Zeolites are an important family of materials with periodic arrays of aluminosilicate cages that are widely used in different industrial processes.
Moreover, they also show interesting electronic phenomena when intercalated with alkali metals\cite{Breck_DW_1973} associated with the electronic states localized within individual cages. 
For example, exotic magnetism with different magnetically ordered states has been reported in alkali-doped zeolites \cite{Nozue_Y_1992, Srdanov_VI_1998, Damjanovic_L_2000, Nakano_T_2006, Nakano_T_2012, Nakano_T_2013}.
With more than 200 possible zeolite frameworks known today\cite{Verheyen_E_2012}, alkali-doped zeolites thus represent a unique playground to control and study the effects of geometry and dopant concentration on the electronic potential depth, electron-electron repulsion and electron-phonon coupling at different length-scales.

\par

Since electronic states are associated to the alkali s-electrons, they remain confined to cages, just like alkali-metals that form clusters or superatoms.
It is thus not surprising that the metallic zeolites have been elusive for many years with only one documented exception in rubidium doped zeolite rho, where the microwave conductivity measurements indicated the metallic ground state\cite{Anderson_2004}. 
Only very recently, the insulator-to-metal transition has been reported in sodium loaded low-silica X (LSX) zeolite, Na$_n$/Na$_{12}$-LSX\cite{Nakano_T_2010,Nozue_Y_2012}.
So far, experimental evidence for the metallic state was mainly limited to the observation of Drude reflection appearing in the infrared region \cite{Nakano_T_2010} and a drastic decrease in the resistivity \cite{Nozue_Y_2012} for the heavily loaded samples.
We stress that the measured resistivity is still very high and atypical of simple metals as it does not decrease with decreasing temperature.
Additional hint of metallic ground state was provided by a precise x-ray diffraction analysis\cite{Ikeda_T_2014}, where it was shown that Na atoms make bonding network through the tunnel windows that connect zeolite cages and thus establish a precondition for a narrow conduction band.
However, firm direct experimental evidence for the metallic state in Na$_n$/Na$_{12}$-LSX is still lacking.

\par

Nuclear magnetic resonance (NMR) is a powerful local-probe experimental tool to investigate a state of matter even in powder and highly air-sensitive samples.
By measuring the temperature dependence of the NMR shift and spin-lattice relaxation time it is in principle possible to distinguish between insulating, metallic and superconducting states \cite{Pennington_1996, Walstedt_2008, Grafe_2008, Potocnik_2014}.
Unfortunately, in alkali-doped zeolites, the spin-lattice relaxation rate, $1/T_1$, is  dominated by strong fluctuations of local magnetic fields and electric field gradients originating from large amplitude atomic motion of alkali metals \cite{Heinmaa_M_2000,Igarashi_M_2013} thus masking the conventional Korringa-like behavior expected in the metallic state.
We show here that at room temperature the values of $^{23}$Na $1/T_1$ due to the Na motion indeed typically exceed by four orders of magnitude contributions from the coupling of nuclear magnetic moments to itinerant electrons in the metallic state. 
Cooling sample to cryogenic temperatures freezes out the atomic motions on the NMR time scale and for Na$_n$/Na$_{12}$-LSX  finally discloses Korringa behavior below 25~K thus confirming the metallic ground state for $n \geq 14.2$.
Surprisingly, a small portion of density of states (DOS) at the Fermi level persists deep into the insulating state.
This important finding that was not possible before with bulk-property measurements, holds important clues about the metal-to-insulator crossover in Na$_n$/Na$_{12}$-LSX, which is here discussed within the correlation-driven and disorder-driven aspects of metal-to-insulator transition (MIT)\cite{Dobrosavljevic_2011,Siegrist_2011}.

%
\section{Experimental}
%
The LSX zeolites have a chemical formula A$_{12}$Al$_{12}$Si$_{12}$O$_{48}$, where A stands for alkali-metal cations that are required for charge compensation of the aluminosilicate framework\cite{Breck_DW_1973}. 
The main structural motif is comprised of truncated octahedral $\beta$ cages, which are arranged in a diamond structure by doubly connecting their 6-membered rings. 
This way additional supercages are formed with a diameter approximately twice of that of $\beta$ cages.
Following the Lowenstein's rule\cite{Loewenstein_W_1954} the Si and Al atoms alternatingly occupy the framework sites resulting in structurally ordered LSX zeolite framework. 

\par

When sodium-based parent structures, i.e. A$=$Na, hereafter abbreviated as Na$_{12}$-LSX, are exposed to Na vapour following the standard procedure described elsewhere\cite{Nozue_Y_2012}, a controlled amount of Na is additionally loaded yielding a targeted composition Na$_n$/Na$_{12}$-LSX. 
Here $n$ denotes the loading density of guest Na atoms per supercage.
The values of $n$ were for the purpose of this study recalibrated by inductively coupled plasma technique.
Particularly for higher density levels, the calibrated values substantially exceed those calculated from known amount of starting materials used in our preliminary NMR study \cite{Igarashi_M_2013}.

\par

Here we present detailed $^{23}$Na NMR experiments on Na$_{n}$/Na$_{12}$-LSX samples in the Na-loading range $11.3 \leq n \leq  16.5$. The $^{23}$Na ($I = 3/2$) NMR spectra and $T_1$ were measured in a magnetic field of 4.7~T in the temperature range between 6~K and 340~K.
The $^{23}$Na reference
frequencies of 52.9055~MHz was determined from NaCl aqueous solution standards. 
Optical reflectance spectra were measured by conventional apparatus.
Since we were dealing with powder samples with a grain diameter of few $\mu$m, the resistivity, $\rho$, was measured by pinching the powder between two metallic plates acting also as terminals\cite{Nozue_Y_2012}.

%
%
\begin{figure} [t]
\includegraphics[width=1.0\linewidth]{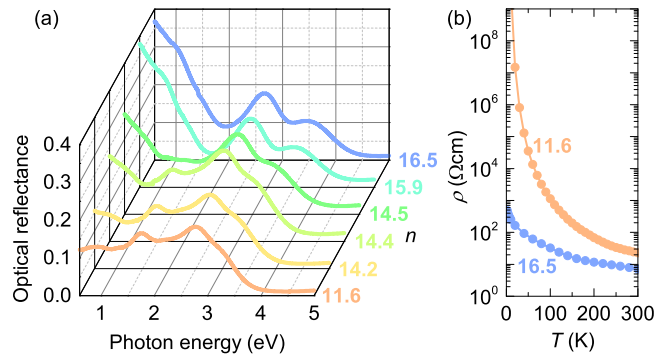}
\caption{(Color online) (a) Room temperature optical conductivity of Na$_{n}$/Na$_{12}$-LSX zeolites as a function of sodium loading level $n$. (b) Temperature dependence of resistivity for insulating ($n=11.6$) and metallic ($n=16.5$) samples.}
\label{fig1}
\end{figure}
%
\section{Experimental Results}
%

The optical reflectance spectra of the samples included in this study clearly demonstrate the emergence of Drude peak at lower photon energies for $n \geq 14.2$ as shown in Fig.~\ref{fig1}(a). 
Moreover, the temperature dependence of resistivity reveals finite low-temperature values for $n = 16.5$, but diverges for $n = 11.6$ as displayed in Fig.~\ref{fig1}(b).  
The measured resistivity may be affected by the constriction resistance at the connection between powder particles, which can result in a misleading negative temperature coefficient of the resistivity for $n = 16.5$.
Nonetheless, finite value of $\rho$ at 2 K for $n = 16.5$ demonstrates a finite DOS at the Fermi level, $N(E_F)$, consistent with the metallic state. These two standard characterization techniques thus comply with the MIT as a function of Na-loading at a critical loading concentration between $n=11.6$ and $n=14.2$, in full agreement with the literature data\cite{Nakano_T_2010,Nozue_Y_2012}.

\par

\begin{figure} [b]
\includegraphics[width=1.0\linewidth]{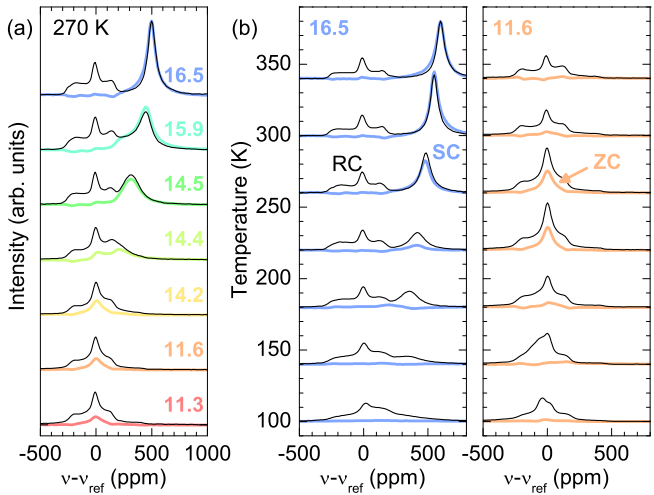}
\caption{(Color online) (a) $^{23}$Na NMR spectra at 270~K as a function of loading level $n$. (b) Temperature dependence of $^{23}$Na spectra for metallic $n=16.5$ and insulating $n=11.6$ samples. All spectra were measured with short (solid black line) and long pulses (thick color line). See text for more details.}
\label{fig2}
\end{figure}

At 270~K, the $^{23}$Na NMR spectrum of insulating Na$_{11.6}$/Na$_{12}$-LSX powder comprises several overlapping peaks close to the Larmor frequency [Fig.~\ref{fig2}(a)]. 
The structure of $^{23}$Na NMR lineshape reflects the multitude of Na sites in the $\beta$ cages and supercages of the LSX structure\cite{Ikeda_T_2014,Feuerstein_M_1998}. 
The bulk magnetic susceptibility of this sample shows a diamagnetic response, although the presence of  diluted localized magnetic moments is revealed by a characteristic low-temperature Curie tail \cite{Nozue_Y_2012}.
Therefore, the predominately diamagnetic susceptibility of $n=11.6$ sample suggests that the lineshape and the shift of the $^{23}$Na NMR spectrum are almost entirely determined by the nuclear chemical shift and quadrupole interactions. 
The insulating Na$_{11.6}$/Na$_{12}$-LSX sample can be set thus as a suitable NMR reference against which all changes of NMR parameters when crossing the MIT are compared.

\par

Indeed, for samples with $n \geq  14.2$, a Lorentzian line [hereafter named as a shifted component (SC)] appears in the metallic samples on the high-frequency side of the $^{23}$Na NMR spectrum, well separated from the diamagnetic frequency range [Fig.~\ref{fig2}(a)]. 
We stress that this line is completely absent in all insulating samples. 
The SC is optimally detected with an echo pulse sequence with precisely two times longer pulse length than that optimized for the residual diamagnetic $^{23}$Na spectral component centered around zero shift -- hereafter we call it a residual component (RC) as it is reminiscent to that described above for the insulating $n=11.6$ sample. 
We conclude that the electric field gradient for Na atoms contributing to this SC is averaged out on the time scale of $^{23}$Na NMR measurements, $\sim 10^{-5}$~s.
Motional effects also explain the Lorentzian lineshape of SC. 
EFG averaged to zero and the Lorentzian lineshape are clear signatures that at elevated temperatures Na atoms undergo large amplitude displacements. 

\par

%
%

The appearance of SC is limited to samples that show metallic-like response in optical and resistivity measurements (Fig.~\ref{fig1}) and is completely absent in insulating samples, e.g. as for $n=11.6$ shown in Fig.~\ref{fig2}(b).
Temperature dependence of $^{23}$Na NMR spectra shown for $n=16.5$ in Fig.~\ref{fig2}(b), reveals that with increasing temperature the intensity of SC increases significantly above $\approx 150$~K. 
This means that such states must be thermally excited from the ground state.
We can rationalize the appearance and the strong temperature dependent shift of SC within a polaron model, where thermally activated behavior is associated with the creation/annihilation of localized small polarons from the bath of (conducting) large polarons\cite{Igarashi_M_2013}.

\par

Unfortunately, the observation of SC for $n\geq 14.2$ is only an indirect proof of metallic state.
Additional complexity in the analysis of $^{23}$Na spectra arises for $n \leq 14.4$, where we observe another non-shifted component (see Fig.~\ref{fig2}), which has the optimal pulse lengths the same as  SC detected at $n \geq 14.2$. 
Using the same arguments as for SC, we conclude that Na atoms contributing to this line must  perform large amplitude jumps between different sites in the cage. 
However, unlike to SC, this line has no hyperfine interaction with unpaired electron spins, therefore we call it zero component(ZC), implying that it has a completely different origin.
It is interesting to note that the $^{23}$Na NMR spectrum for $n = 14.4$ [Fig.~\ref{fig2}(a)] shows a coexistence of ZC and SC, which may indicate an inhomogeneous distribution of sodium atoms throughout the zeolite cages. 

\par

%
%
\begin{figure} [t!]
\includegraphics[width=1.0\linewidth]{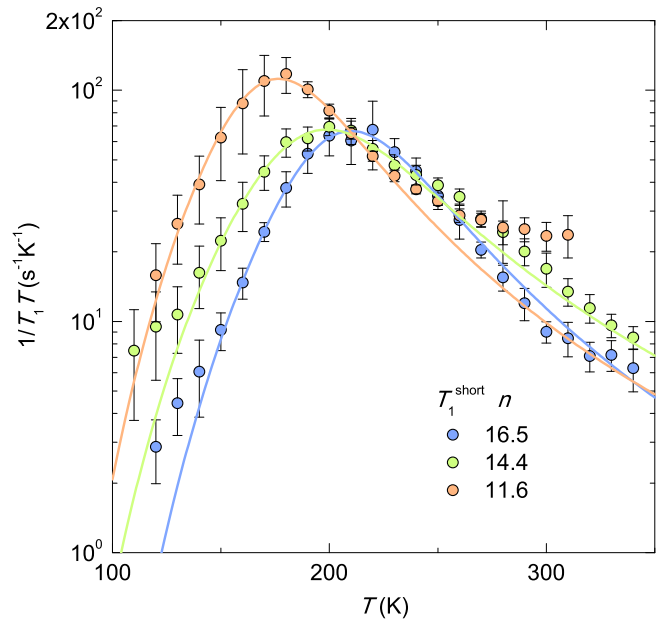}
\caption{(Color online) Representative temperature dependences of $^{23}$Na $1/T_1T$ for the SC/ZC component with shorter relaxation time.
The solid lines are fits obtained from the BPP model described by Eq.~(\ref{BPP}).}   
\label{fig3}
\end{figure}

Since the metallic state cannot be unambiguously confirmed from the analysis of $^{23}$Na NMR spectra, we next move to spin-lattice relaxation data, where the Korringa behavior ($1/T_1T = {\rm const.}$) is a signature of metallic state\cite{Walstedt_2008,Slichter_1989}.
For all compositions RC has, compared to ZC and SC, by two orders of magnitude longer $T_1$.
This distinction is manifested over a wide temperature range as two separate nuclear magnetization recoveries with long and short time constants. 
The temperature dependence of the short ZC/SC $T_1$ component (Fig.~\ref{fig3}), has a pronounced maximum in $1/T_1 T$, which can be between 100 and 350~K empirically modeled within the Bloembergen-Purcell-Pound (BPP)-type mechanism\cite{BPP_1948}
\begin{equation}
\label{BPP} 
\left( \frac{1}{T_{1}T} \right)_{\rm BPP}=\frac{C}{T} \frac{\tau_c}{1+\omega^2\tau_c^2}. 
\end{equation}
Here $\tau_c$ is the correlation time for the local field fluctuations at the nucleus, $\omega$ is the Larmor angular frequency and $C$ is a measure of the fluctuating local fields magnitude. 
We estimated the activation energy for the local field fluctuations by assuming the Arrhenius type correlation time and obtained a value of around $0.1$~eV for all loading densities investigated.  
The estimated activation energy is typical for atomic motion and provides yet another independent proof that sodium motion is present in both insulating and metallic state.
However, such strong relaxation due to the atomic motion completely masks the metallic Korringa contribution to the spin-lattice relaxation.

\par

%
%
\begin{figure} [b!]
\includegraphics[width=1.0\linewidth]{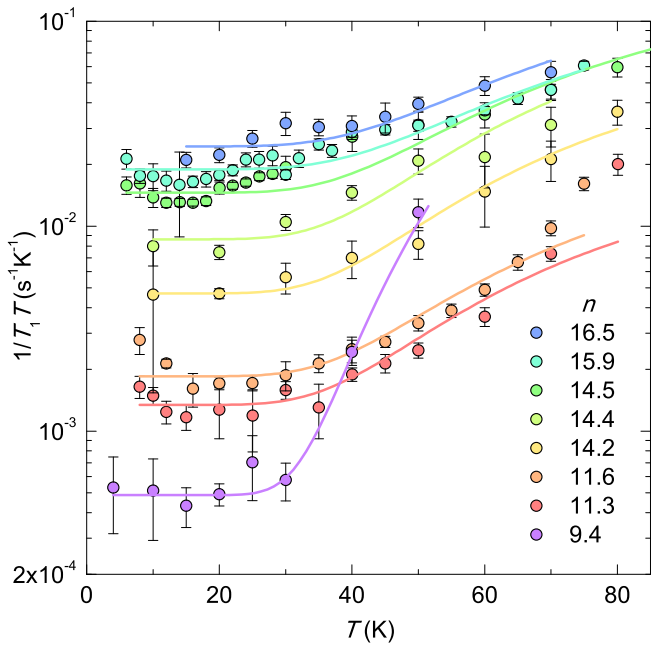}
\caption{(Color online) Low-temperature dependences of $^{23}$Na $1/T_1T$ showing plateau-like behavior below 25~K for various loading levels $n$.
The solids lines are fits obtained by combining the BPP and Korringa relaxation mechanisms [Eqs.~(\ref{total})].
For comparison we added the low-temperature data for the $n=9.4$ sample, which is according to the optical reflectance and resistivity measurements considered to lie deep in the insulating phase.}
\label{fig4}
\end{figure}

Since $\tau_{\rm c}$  increases exponentially with decreasing temperature, we anticipate that according to Eq. (1) the BPP contribution to the total $^{23}$Na relaxation rate diminishes at low temperatures, i.e., the atomic-motion driven $(1/T_1T)_{\rm BPP}\rightarrow 0$ as temperature decreases.
In addition, the relaxation rates for RC and ZC/SC become comparable below $\sim 100$~K thus implying that other relaxation mechanisms, which are only moderately site dependent, set in.
Therefore, the low temperature relaxation curves were fitted with stretched exponential model
$\sim {\rm exp} [-(\tau /T_1)^\alpha]$ with a single effective $T_1$, where a factor $\alpha \approx 0.6$ accounts for distribution of $T_1$'s due to multiple sites.
The most important observation arising from this analysis  is that plotting $1/T_1T$ versus temperature for metallic samples with $n \geq 14.2$, we finally find temperature independent $1/T_1T$ below $\sim 25$~K  (Fig.~\ref{fig4}).
This is a hallmark of the nuclear spin-lattice relaxation via the itinerant electrons and is accounted for by the Korringa expression \cite{Walstedt_2008,Slichter_1989}
\begin{equation}
\label{Korringa}
\left( \frac{1}{T_{1}T} \right)_{\rm metal} = \frac{4\pi k_B}{\hbar}\frac{\gamma_{\rm Na}^2}{\gamma_e^2}K_{\rm iso}^2 \propto [N(E_F)]^2.
\end{equation}
Here $\gamma_e$ and $\gamma_{\rm Na}$ are the electronic and $^{23}$Na gyromagnetic ratios, respectively.
The temperature independent isotropic Knight shift, $K_{\rm iso}$, which is proportional to Pauli spin susceptibility, is a measure of DOS at the Fermi level, $N(E_F)$. 
Comparing the low-temperature $(1/T_1T)$ values for different $n$, we find that in metallic samples $1/T_1T$ is enhanced relative to insulating samples by an order of magnitude thus corroborating with the additional relaxation channel presented by Korringa relaxation mechanism.
Moreover, a monotonic increase of low-temperature $(1/T_1T)$ with $n$ for $n \geq 14.2$ speaks for a monotonic increase of $N(E_F)$ with Na loading level.

\par

In order to quantitatively extract $(1/T_1T)_{\rm metal}$ from the measured temperature dependence of $1/T_1T$, we assumed that the relaxation rate has two contributions, the BPP and Korringa, described by Eqs.~(\ref{BPP}) and (\ref{Korringa}), respectively. That is, the data was fitted to
\begin{equation}
\label{total}
\left( \frac{1}{T_{1}T} \right) = \left( \frac{1}{T_{1}T} \right)_{\rm BPP}+\left( \frac{1}{T_{1}T} \right)_{\rm metal}.
\end{equation}
Table~\ref{DOS} summarizes the values of  $\left( 1/T_{1}T \right)_{\rm metal}$ for all $n$ in the range between 9.4 and 16.5 and is compared to the related value measured in bulk metallic Na. 
We note that the $(1/T_1T)_{\rm metal}$ values are by more than one order of magnitude smaller than that of bulk metallic Na thus indicating a relatively low $N(E_F)$ in these metallic samples. 
This conclusion is further supported, if we use $(1/T_1T)_{\rm metal} = 2.3\times 10^{-2}$~s$^{-1}$K$^{-1}$ of $n = 16.5$ sample to calculate $K_{\rm iso} = 300$~ppm from Eq.~(\ref{Korringa}) and compare it to much larger value of 1120~ppm found in metallic sodium \cite{Walstedt_2008}.
Surprisingly, the plateau-like low-temperature behavior in $1/T_1T$ is also seen for $n=11.6$ and $11.3$, where the Drude term is not observed and resistivity diverges at low temperature.
We stress that the observed low-temperature $1/T_1T$ clearly rules out the possibility that for this loading range the system can be discussed as a narrow-gap semiconductor. In this case, $1/T_1T$ would be dominated by the $\exp(-\Delta/k_BT)$ term for $k_BT \ll \Delta$, which decays exponentially with decreasing temperature\cite{Grykalowska_2007}, in disagreement with the experimental data.

\begin{table}
\caption{Extracted values of $^{23}$Na $\left( 1/T_{1}T \right)_{\rm metal}$ as a function of Na loading level $n$. The values of $N(E_F)$ normalized to metallic sodium $^{\rm Na}N(E_F)$ listed in column three are calculated using Eq.~(\ref{Korringa}) and taking $\left( 1/T_{1}T \right)=0.210 \pm 0.004$~s$^{-1}$K$^{-1}$ of metallic sodium\cite{Walstedt_2008}.}
\begin{tabular}{cccc}
\hline \hline
Loading & $\left( 1/T_{1}T \right)_{\rm metal}$ & $N(E_F)/^{\rm Na}N(E_F)$  
\\
level $n$ & (s$^{-1}$K$^{-1}$) &
\\
\hline 
$16.5$ & $0.0245 \pm 0.0017$ & $0.343 \pm 0.031$  
\\ 
$15.9$ & $0.0190 \pm 0.0006$ & $0.302 \pm 0.016$
\\
$14.5$ & $0.0146 \pm 0.0004$ & $0.264 \pm 0.012$
\\
$14.4$ & $0.0086 \pm 0.0008$ & $0.204 \pm 0.023$
\\
$14.2$ & $0.0047 \pm 0.0004$ & $0.150 \pm 0.017$
\\
$11.6$ & $0.0019 \pm 0.0001$ & $0.094 \pm 0.007$
\\
$11.3$ & $0.0013 \pm 0.0001$ & $0.080 \pm 0.008$
\\
$9.4$ & $0.0005 \pm 0.0001$ & $0.049 \pm 0.005$
\\ 
\hline \hline
\end{tabular}
\label{DOS}
\end{table}

%
\section{Discussion}
\begin{figure} [t!]
\includegraphics[width=1.0\linewidth]{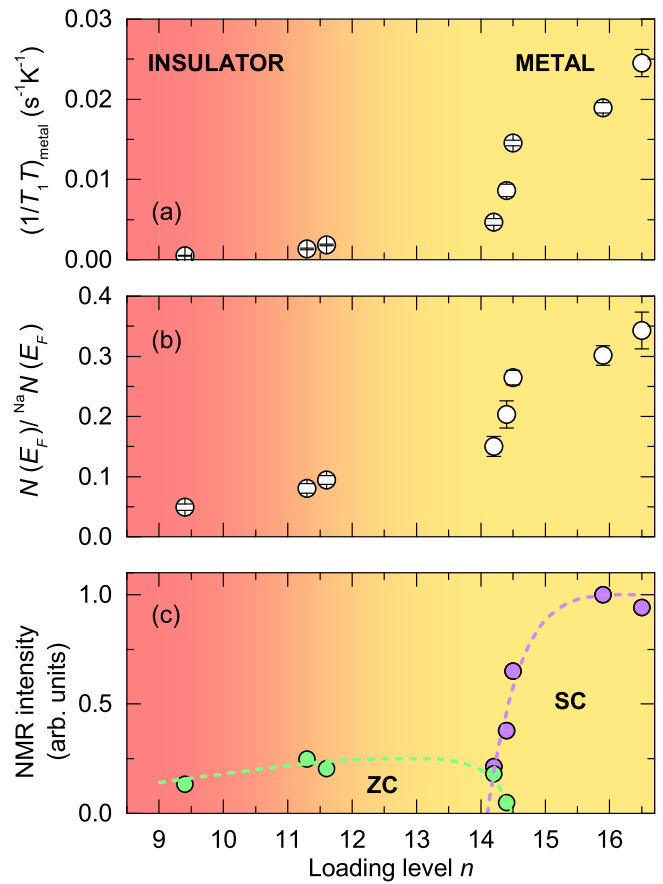}
\caption{(Color online) The phase diagram of Na$_n$/Na$_{12}$-LSX zeolite showing the values of (a) $\left( 1/T_{1}T \right)_{\rm metal}$, (b) normalized $N(E_F)$ and (c) NMR spectrum intensity for ZC and SC as a function of Na loading level $n$. 
The color gradient divides the insulating and metallic regions as experimentally observed from the resistivity and optical conductivity measurements.}
\label{fig5}
\end{figure}

%
Plotting the normalized $N(E_F)$ extracted from Eq.~(\ref{Korringa}) versus the sodium loading level $n$ (Fig.~\ref{fig5}), we find that $N(E_F)$ markedly increases with $n$ for $n \geq  14.2$, thus speaking for the enhancement of DOS at Fermi level $N(E_F)$ with doping, which is in qualitative agreement with the enhancement of optical reflectance [Fig.~\ref{fig1}(a)]. 
We note that for the most loaded sample ($n = 16.5$), the extracted $N(E_F)$ is by a factor of $\sim 3$ smaller than the corresponding value in bulk Na.
Our study of a minimal single-orbital Hubbard model revealed\cite{Zitko_2015} that the experimentally observed strong variation of Drude peak in optical conductivity cannot be explained solely by band filling effects. In fact, it shows that the variation of electron-electron repulsion $U$ divided by a bandwidth $W$, is much more relevant, meaning that the main effect of sodium loading is to define the electronic potential and the Coulomb repulsion felt by the electrons in the zeolite cages.
Similarly, the importance of electron correlations has been theoretically\cite{Arita_2004} and experimentally\cite{Nakano_1999,Ikemoto_2000} recognized for related potassium-loaded LTA and FAU zeolites.
\par
%
%
%

However, the correlation-driven MIT is expected to be of first order\cite{Dobrosavljevic_2011}, which is not directly supported by the present data.
We recall that a finite $N(E_{\rm F})$ has been observed even in the nominally insulating samples of Na$_n$/Na$_{12}$-LSX.
For example, the $n=11.3$ sample exhibits a resistivity diverging at low temperatures and is described by an energy band gap of 0.2~eV.
At the same time, the local probe $^{23}$Na NMR for this sample shows the finite $N(E_{\rm F})$, more precisely $\sim 8$\% of the value found in metallic sodium.
The phase diagram shown in Fig.~\ref{fig5} is  more reminiscent of a metal-to-insulator crossover rather than the sharp transition thus calling for considerations of other relevant microscopic factors.
\par
In the alternative picture, where disorder is the driving mechanism for MIT, a continuous metal-to-insulator is typically found at finite temperatures. \cite{Rosenbaum_1980}
The reason is that at finite temperatures the electrons can escape the trapping potential through the thermal activation.
Indeed, the alkali-doped zeolites can be viewed as a strongly disordered system for several reasons.
First, by a careful analysis of the spectral intensities belonging to SC and ZC as a function of sodium loading level [Fig.~\ref{fig5}(c)], we identify a region around $n=14.4$ where both spectral components coexist, implying an inhomogeneous Na distribution in the cages that is responsible for a cage-to-cage variation in the electric potential depths. 
Second, as pointed out in the x-ray study by Ikeda {\it et al.}\cite{Ikeda_T_2014}, in Na$_n$/Na$_{12}$-LSX the coordinates of available sodium sites in the zeolite framework remain the same as a function of sodium loading level.
What is changing with $n$ is the average occupancy of these sites. Local variations in the Na arrangements in the neighboring cages is responsible for the variations in the local potential depth and thus for the disorder.
For low carrier densities, the trapping potential of disordered sodium clusters within the cages is expected to become comparable or larger than the Fermi level, and the electrons get localized.
At higher loading densities not only the carrier density increases, but also the disorder strength decreases since the loaded sodium atoms reach the limit of the highest possible occupancy in the cages\cite{Ikeda_T_2014}, explaining the MIT.
\par
However, at the critical loading densities yet another possibility of a percolation-type metal-to-insulator transition\cite{Dobrosavljevic_2011} opens.
Although, the disorder-driven scenario in the presence of varying correlation effects seems plausible, we should be aware that the variation of alkali atom loading level not only changes the amount of disorder but also strongly determines the electric potential felt by the electrons.
Using electron-density distribution analysis Ikeda {\it et al.}\cite{Ikeda_T_2014} observed a formation of the chain-like Na cation distribution in metallic Na$_{16.7}$/Na$_{12}$-LSX, which connects the neighboring supercages.
This Na-Na connectivity is not formed in nominally insulating Na$_{9.4}$/Na$_{12}$-LSX.
In the percolation picture of a random (disordered) potential a small metallic regions of connected supercages are formed at low sodium loading densities separated by insulating areas.
When the electron density increases the metallic regions grow and eventually become connected at the percolation threshold\cite{Dobrosavljevic_2011}.
Although our observation of the crossover regime and the inhomogeneous distribution of sodium atoms throughout the zeolite lattice tentatively support the percolation picture, we leave the important question of the true nature of MIT in alkali doped zeolites open for further studies.
However, what is made clear from present results is that for the Na$_n$/Na$_{12}$-LSX zeolites the concentration of sodium atoms strongly affects in a very complex way the electronic properties by simultaneously changing the band filling, the electronic potential depth, the electron-electron repulsion and the amount of disorder.
\par
%
%
\section{Summary}
%
Using NMR as a local probe of sodium loaded low-silica X zeolite (Na$_n$/Na$_{12}$-LSX), we have unambiguously confirmed a metallic ground state for higher loading densities of $n \geq 14.2$.
By extracting the DOS at the Fermi level as a function of sodium loading level, we have shown a rather continuous (crossover like) evolution across the metal-to-insulator transition.
Most importantly, a finite DOS at the Fermi level for nominally insulating samples and a clear indication of inhomogeneous Na distribution in the neighboring cages in the crossover region, put some constraints on the driving mechanism of electron localization and the nature of MIT in alkali-doped zeolites.

%
\section{Acknowledgment}
%
We thank Dr. T. Ikeda for helpful discussions about the structure of Na$_n$/Na$_{12}$-LSX zeolite. 
MI especially thanks to Dr. T. Shimizu, Dr. A. Goto, Dr. K. Hashi, and Mr. S. Ohki.
This study was partially supported by Grants-in-Aid for Scientific Research (KAKENHI) [Grants No. 24244059(A) and No.26400334(C)] from Japan Society for the Promotion of Science.

\end{document}